\documentclass[smallextended]{svjour3} 

\usepackage{amsmath,amsfonts}
\usepackage{graphicx}
\usepackage{subfigure}

\usepackage{comment}
\DeclareMathOperator{\tr}{tr}
\newcommand{\Tr}{\tr}
\newcommand{\dTr}{d_{\tr}}
\newcommand{\dE}{d_{E}}
\newcommand{\dTrK}{\overline{d}_{\tr}}
\newcommand{\DF}{f_\mathrm{DF}}
\newcommand{\QDF}{f_\mathrm{QDF}}

\newcommand{\ie}{\emph{i.e.\/}}

\begin{document}

\title{Pattern recognition on the quantum Bloch sphere}

\author{Giuseppe Sergioli \and
        Enrica Santucci \and
        Luca Didaci \and
        Jaros{\l}aw A. Miszczak \and
        Roberto Giuntini        
}

\institute{Giuseppe Sergioli \at 
    Universit\`a di Cagliari,
    Via Is Mirrionis 1, I-09123 Cagliari, Italy.\\
    \email{giuseppe.sergioli@gmail.com}
    \and
    Enrica Santucci \at 
    Universit\`a di Cagliari,
    Via Is Mirrionis 1, I-09123 Cagliari, Italy.\\
    \email{enrica.santucci@gmail.com}
    \and
    Luca Didaci \at 
    Universit\`a di Cagliari,
    Via Is Mirrionis 1, I-09123 Cagliari, Italy.\\
    \email{didaci@diee.unica.it}
    \and
    Jaros{\l}aw A. Miszczak  \at 
    Institute of Theoretical and Applied Informatics, Polish Academy
of Sciences,\\ Ba{\l}tycka 5, 44-100 Gliwice, Poland\\
Universit\`a di Cagliari,
    Via Is Mirrionis 1, I-09123 Cagliari, Italy.\\
    \email{miszczak@iitis.pl}
    \and
    Roberto Giuntini \at 
    Universit\`a di Cagliari,
    Via Is Mirrionis 1, I-09123 Cagliari, Italy.\\
    \email{giuntini@unica.it}
    }

\date{Received: date / Accepted: date}

%\title[Pattern recognition framework based on the Bloch \dots]%
%	{Pattern recognition on the quantum Bloch sphere}

%\author[Sergioli]{Giuseppe Sergioli}
%\address[G.~Sergioli]{
%    Universit\`a di Cagliari,
%    Via Is Mirrionis 1, I-09123 Cagliari, Italy.}
%    \email{giuseppe.sergioli@gmail.com}

%\author[Santucci]{Enrica Santucci}
%\address[E.~Santucci]{
%    Universit\`a di Cagliari,
%    Via Is Mirrionis 1, I-09123 Cagliari, Italy.}
%    \email{enrica.santucci@gmail.com}

%\author[Didaci]{Luca Didaci}
%\address[L.~Didaci]{
%Universit\`a di Cagliari,
%  Via Is Mirrionis 1,
%I-09123 Cagliari, Italy} \email{didaci@unica.it}

%\author[Miszczak]{Jaros{\l}aw A. Miszczak}
%\address[J.A.~Miszczak]{Institute of Theoretical and Applied Informatics, Polish Academy
%of Sciences, Ba{\l}tycka 5, 44-100 Gliwice, Poland \and
%    Universit\`a di Cagliari,
%    Via Is Mirrionis 1, I-09123 Cagliari, Italy.}
% \email{jmiszczak@acm.org}

%\author[Giuntini]{Roberto Giuntini}
%\address[R.~Giuntini]{
%    Universit\`a di Cagliari,
%    Via Is Mirrionis 1, I-09123 Cagliari, Italy.}
%    \email{giuntini@unica.it}

\maketitle

\begin{abstract}
We introduce a framework suitable for describing pattern recognition task using
the mathematical language of density matrices. In particular, we provide a
one-to-one correspondence between patterns and pure density operators. This correspondence enables us to: $i)$ represent the Nearest Mean Classifier (NMC) in terms of quantum objects, $ii$) introduce a Quantum Classifier (QC). By comparing the QC with the NMC on different 2D datasets, we show the first classifier can provide additional information that are particularly beneficial on a classical computer with respect to the second classifier.   
%The main advantage of the introduced framework is that it provides an
%uniform method for describing the uncertainty of the data. 

\keywords{Bloch sphere \and pattern recognition \and geometry of quantum states}
 \PACS{03.67.Ac \and 03.65.Aa}
\end{abstract}

%%%%%%%%%%%%%%%%%%%%%%%%%%%%%%%%%%%%%%%%%%%%%%%%%%%%%%%%%%%%%%%%%%%%%%%%%%%%%%%%
\section{Introduction}
%%%%%%%%%%%%%%%%%%%%%%%%%%%%%%%%%%%%%%%%%%%%%%%%%%%%%%%%%%%%%%%%%%%%%%%%%%%%%%%%

Quantum machine learning aims at merging the methods from quantum information
processing and pattern recognition to provide new solutions for problems in the
areas of pattern recognition and image
understanding~\cite{schuld14introduction,wittek14quantum,wiebe15quantum}. In the
first aspect the research in this area is focused on the application of the
methods of quantum information processing~\cite{miszczak12high-level} for
solving problems related to classification and clustering~\cite{trugenberger2002quantum,caraiman2012image}. One of the possible
directions in this field is to provide a representation of computational models
using quantum mechanical concepts. From the other perspective the methods for
classification developed in computer engineering are used to find solutions for
problems like quantum state
discrimination~\cite{helstrom,chefles00quantum,hayashi05quantum,lu2014quantum},
which ares tightly connected with the recent developments in quantum
cryptography.
\\Using quantum states for the purpose of representing patterns is naturally
motivated by the possibility to exploit quantum algorithms to boost the
computational intensive parts of the classification process. In particular, it
has been demonstrated that quantum algorithms can be used to improve the time
complexity of the $k-$\emph{nearest neighbor} ($k$NN) method. Using the
algorithms presented in \cite{wiebe15quantum} it is possible to obtain
polynomial reductions in query complexity in comparison to the corresponding
classical algorithm. 

Another motivation comes from the possibility of using quantum-inspired
algorithms for the purpose of solving classical problems. Such an approach has been
exploited by various authors. In \cite{tanaka08quantum} authors propose an
extension of Gaussian mixture models by using the statistical mechanics point of
view. In their approach the probability density functions of conventional
Gaussian mixture models are expressed by using density matrix representations.
On the other hand, in \cite{ostaszewski15quantum} authors utilize the
quantum representation of images to construct measurements used for
classification. Such approach might be particularly useful for the
physical implementation of the classification procedure on quantum machines.

In the last few years, many efforts to apply the quantum formalism to
non-microscopic contexts \cite{AeDho,AGS,EWL,N,Ohya,Schwartz,Stapp} and to
signal processing \cite{Eldar} have been made. Moreover, some attempts to connect
quantum information to pattern recognition can be found in
\cite{schuld14introduction,schuld14quantum,schuld14quest}. Exhaustive survey and
bibliography of the developments concerning applications of quantum computing in
computational intelligence are provided in
\cite{manju14applications,wittek14quantum}. Even if these results seem to
suggest some possible computational advantages of an approach of this sort, an
extensive and universally recognized treatment of the topic is still missing
\cite{schuld14introduction,lloyd2014quantum,lloyd13quantum}.

The main contribution of our work is the introduction of a new framework
to encode the classification process by means of the mathematical language of density
matrices~\cite{beltrametti2014quantum-p1,beltrametti2014quantum-p2}. We show that this representation leads to two different developments: $i)$ it enables us to provide a representation of the \emph{Nearest Mean Classifier} (NMC) in terms of quantum objects; $ii)$ it can be used to introduce a \emph{Quantum Classifier} (QC) that can provide a significative improvement of the performances on a classical computer with respect to the NMC.   
%This paper is organized as follows. The remaining part of this section provides
%required preliminaries and discussed the previous attempts to tackle the problem
%of pattern recognition in the quantum computation framework. In
%Section~\ref{sec:density-patterns} we develop the framework where density
%matrices are used to describe the patters. For the case without the uncertainty
%the presented framework can be used to reproduce the classical procedure. 

The paper is organized as follows. In Section \ref{sec:pre} basic notions of
quantum information and pattern recognition are introduced. In Section
\ref{sec:pdp} we formalize the correspondence between arbitrary two-feature
patterns and pure density operators and we define  the notion of \emph{density
pattern}. In Section \ref{sec:cp} we provide a representation of NMC by using density patterns and by the introduction of an \emph{ad hoc} definition of distance between quantum states. In Section \ref{qcp} is devoted to describe a new quantum classifier QC that has not a classical counterpart in the standard classification process.  Numerical simulations for both QC and NMC are presented. In Section \ref{sec:general} a geometrical idea to
generalize the model to arbitrary $n$-feature patterns is proposed. Finally, in
Section \ref{sec:concl} concluding remarks and further
developments are discussed. 

%%%%%%%%%%%%%%%%%%%%%%%%%%%%%%%%%%%%%%%%%%%%%%%%%%%%%%%%%%%%%%%%%%%%%%%%%%%%%%%%
\section{Representing classical and quantum information quantities}\label{sec:pre}
%%%%%%%%%%%%%%%%%%%%%%%%%%%%%%%%%%%%%%%%%%%%%%%%%%%%%%%%%%%%%%%%%%%%%%%%%%%%%%%%

%%%%%%%%%%%%%%%%%%%%%%%%%%%%%%%%%%%%%%%%%%%%%%%%%%%%%%%%%%%%%%%%%%%%%%%%%%%%%%%%

%%%%%%%%%%%%%%%%%%%%%%%%%%%%%%%%%%%%%%%%%%%%%%%%%%%%%%%%%%%%%%%%%%%%%%%%%%%%%%%%
In the standard quantum information theory \cite{BenShor,Shann}, the states of
physical systems are described by unit vectors and
their evolution is expressed in term of unitary matrices (\ie\ quantum gates).
However, this representation can be applied for an ideal case only, because it
does not take into account some unavoidable physical phenomena, such as
interactions with the environment and irreversible transformations. In modern
quantum information theory \cite{Jaeg,Jaeg2,Wilde}, another approach is
adopted. The states of physical systems are described by density operators --
also called \emph{mixed states} \cite{AKN,DGG,FSA} -- and their evolution is
described by quantum operations. The space $\Omega_n$ of density operators for
$n$-dimensional system consists of positive semidefinite matrices with unit
trace. 

A quantum state can be \emph{pure} or \emph{mixed}. We say that a state of a
physical system is pure if it represents ``maximal'' information about the
system, \ie\ an information that can not be improved by further observations.
A probabilistic mixture of pure states is said to be a \emph{mixed} state.
Generally, both pure and mixed states are represented by density operators, that are positive and Hermitian
operators (with unitary trace) living in a $n$-dimensional complex Hilbert space
$\mathcal H$. 
Formally, a density operator $\rho$ is pure iff
$\tr(\rho)^2=1$ and it is mixed iff $\tr(\rho)^2<1$.
%
%Each density operator on
%the Hilbert space $\otimes^n\mathbb C^2$ can be represented as a convex
%combination of \emph{pure states}:
%$\rho=\sum_i\lambda_i\ket{\psi_i}\bra{\psi_i},\hspace{0.2cm}\sum_i\lambda_i=1,
%\ket{\psi_i}\in\otimes^n\mathbb C^2$.
\\If we confine ourselves in the $2$-dimensional Hilbert space $\mathcal H$, a
suitable representation of an arbitrary density operator $\rho\in\Omega_2$ is
provided by 
\begin{equation}\label{Bl}
\begin{split}
\rho&=\frac{1}{2}(I+r_1\sigma_1+r_2\sigma_2+r_3\sigma_3)=\\
&=\frac{1}{2}
\begin{pmatrix}1+r_3 & r_1-ir_2 \\
          r_1+ir_2 & 1-r_3  
\end{pmatrix},
\end{split}
\end{equation}
where $\sigma_i$ are the Pauli matrices. This expression comes to be useful in
order to provide a geometrical representation of $\rho$. Indeed, each density
operator $\rho\in\Omega_2$ can be geometrically represented as a point of a
radius-one sphere centered in the origin (the so called \emph{Bloch sphere}),
whose coordinates (\ie\ \emph{Pauli components}) are $r_i$ (with
$\sum_ir_i^2\leq1$). By using the generalized Pauli matrices \cite{Bertl,kimura03bloch}
it is also possible to provide a geometrical representation for an arbitrary
$n$-dimensional density operator, as it will be showed in Section
\ref{sec:general}.
%For any density operator $\rho\in\Omega_n$ it is possible to define a
%\emph{measure of mixedness}, given by the \emph{normalized linear entropy}
%\cite{Peters}
%\begin{equation}\label{le}
%S_L(\rho)=\frac{2^n}{2^n-1}(1-\Tr(\rho)^2).
%\end{equation}
%
%By a straightforward calculation, one can verify that $S_L(\rho)=0$ iff $\rho$
%is pure and $S_L(\rho)=1$ iff $\rho$ is \emph{maximally} mixed. 
Again, by
restricting to a $2$-dimensional Hilbert space, the points on the surface of the
Bloch sphere represent pure states, while the inner points represent mixed
states.

Quantum formalism turns out to be very useful not only in the microscopic scenario but also to encode classical data. This has naturally suggested several attemps to represent the standard framework of machine learning through the quantum formalism \cite{lloyd13quantum,schuld14introduction}.
In particular, pattern recognition \cite{webb,DuHa} is the scientific discipline which deals with theories and methodologies for designing algorithms and machines capable of automatically recognizing ``objects''  (i.e. patterns) in noisy environments.
% \footnote{Quoting Duin et al. ``Pattern recognition can be defined as the scientific discipline that studies theories and methods for designing machines able to recognise patterns in noisy data''\cite{Duin}.}
Some typical applications are multimedia document classification, remote-sensing image classification, people identification
using biometrics traits as fingerprints. 
%In a real environment, the process of classification can individuate the observed object not with certainty but with some probability only. This may be due to intrinsic limitations of the representation space of the objects, or  to the fact that the noise causes a loss of information. This
%suggests to tackle the pattern recognition problems with the quantum framework.
%It has a probabilistic structure and thus allows the natural treatment of the
%uncertanity.

%%%%%%%%%%%%%%%%%%%%%%%%%%%%%%%%%%%%%%%%%%%%%%%%%%%%%%%%%%%%%%%%%%%%%%%%%%%%%%%%
%%%%%%%%%%%%%%%%%%%%%%%%%%%%%%%%%%%%%%%%%%%%%%%%%%%%%%%%%%%%%%%%%%%%%%%%%%%%%%%%
A pattern is a representation of an object. The object could be concrete (i.e., an animal, and the pattern recognition task could be to identify the kind of animal) or an abstract one (i.e. a facial expression, and the task could be to identify the emotion expressed by the  facial expression). The pattern is characterized via a set of measurements called \emph{features}\footnote{Hence, as a pattern is an object characterized by the knowledge of its
features, analogously, in quantum mechanics a state of a physical system is
represented by a density operator, characterized by the knowledge of its
observables.}. Features can assume the forms of categories, structures, names, graphs, or, most commonly, a vector of real number (feature vector) $\mathbf{ x}  = (x_1, x_2, \dots, x_d)\ \in \mathbb{R}^d $.  
Intuitively, a class is the set of all similar patterns. For the sake of simplicity, and without loss of generality, we assume  that each object belongs to one and only one class, and we will limit our attention to 2-class problems. For example, in the domain of `cats and dogs' we can consider the classes $C_{cats}$ (the class of all cats) and $C_{dogs}$ (the class of all dogs). The pattern at hand is either a cat or a dog, and a possible representation of the pattern could consist in the height of the pet and the length of its tail. In this way, the feature vector $\mathbf{ x_1} = (x_{11},x_{12}) $  is the pattern representing a pet whose height and length of the tail are $x_{11}$ and $x_{12},$ respectively.
\\Now, let us consider an object $\mathbf{ x}_t$ whose membership class is unknown. The basic aim of the classification process is to establish which class  $\mathbf{ x}_t$  belongs to. To reach this goal, standard pattern recognition designs a  \emph{classifier} that, given the feature vector $\mathbf{ x}_t$, has to determine the true class of the pattern. The classifier should take into account all the available information about the task at hand (i.e., information about the statistical distributions of the patterns and information obtained from a set of patterns whose true class is known). This set of patterns is called `training set', and it will be used to define the behavior of the classifier.
\\If no information about the statistical distributions of the patterns  is available, an easy classification algorithm that could be used is the \emph{Nearest Mean Classifier} (NMC) \cite{manning2008introduction,HaTi}, or minimum distance classifier. 
The NMC
\begin{itemize}
\item computes the centroids of each class, using the patterns on the training set
 $\mu^*_i = \frac{1}{n_i}\sum_{   \mathbf{ x}   \in C_i} \mathbf{ x} $ where $n_i$ is the number of patterns of the training set belonging to the class $C_i$;
\item assigns the unknown pattern $\mathbf{ x}_t$ to the class with the closest centroid.
\end{itemize}

In the next Section we provide a representation of arbitrary 2D patterns by means of density matrices, while in Section \ref{sec:cp} we introduce a representation of NMC in terms of quantum objects.

%%%%%%%%%%%%%%%%%%%%%%%%%%%%%%%%%%%%%%%%%%%%%%%%%%%%%%%%%%%%%%%%%%%%%%%%%%%%%%%%
\section{Representation of $2$-dimensional patterns}\label{sec:pdp}
%%%%%%%%%%%%%%%%%%%%%%%%%%%%%%%%%%%%%%%%%%%%%%%%%%%%%%%%%%%%%%%%%%%%%%%%%%%%%%%%
Let $\mathbf{ x_i}=(x_{i1},\ldots,x_{ik})$ be a generic pattern, \ie\ a point in $\mathbb
R^k$. By means of this representation, we consider all the $k$ features of $\mathbf{ x_i}$
as perfectly known. Therefore, $\mathbf{ x_i}$ represents a maximal kind of information,
and its natural quantum counterpart is provided by a pure state. For the sake of
simplicity, we will confine ourselves to an arbitary two-feature pattern
indicated by $\mathbf{ x}=(x,y) \footnote{In the standard pattern recognition theory, the symbol $y$ is generally used to identify the label of the pattern. In this paper, for the sake of semplicity, we agree with a different notation.}.$ In this section, a one-to-one correspondence between
each pattern and its corresponding pure density operator is provided.
\\The pattern $\mathbf{ x}$ can be represented as a point in $\mathbb R^2$. The
stereographic projection \cite{Cox} allows to unequivocally map any point
$r=(r_1,r_2,r_3)$ of the surface of a radius-one sphere $\mathbb S^2$ (except
for the north pole) onto a point $\mathbf{ x}=(x,y)$ of $\mathbb R^2$ as
\begin{equation}\label{sp}
SP:(r_1,r_2,r_3) \mapsto \left(\frac{r_1}{1-r_3},\frac{r_2}{1-r_3}\right).
\end{equation}
The inverse of the stereographic projection is given by
\begin{equation}\label{sp1}
SP^{-1}:(x,y) \mapsto \left(\frac{2x}{x^2+y^2+1},\frac{2y}{x^2+y^2+1},\frac{x^2+y^2-1}{x^2+y^2+1}\right).
\end{equation}
%By considering the geometrical representation of a density operator we have
%introduced before, and we put
%\begin{equation}
%\r_1=\frac{2x}{x^2+y^2+1}$
%\item $r_2=\frac{2y}{x^2+y^2+1}$
%\item $r_3=\frac{x^2+y^2-1}{x^2+y^2+1}$
%\end{itemize}
Therefore, by using the Bloch representation given by Eq.~(\ref{Bl}) and placing
\begin{equation}\label{r123}
r_1=\frac{2x}{x^2+y^2+1},\quad  
r_2=\frac{2y}{x^2+y^2+1},\quad 
r_3=\frac{x^2+y^2-1}{x^2+y^2+1},
\end{equation} 
we obtain the following definition.

\begin{definition}[Density Pattern]\label{dp}
Given an arbitrary pattern $\mathbf{ x}=(x,y)$, the density pattern (DP) $\rho_\mathbf{ x}$
associated to $\mathbf{ x}$ is the following pure density operator
\begin{equation}\label{eqn:desnity-pattern}
\rho_\mathbf{ x}=\frac{1}{2}\begin{pmatrix}1+r_3 & r_1-ir_2 \\
          r_1+ir_2 & 1-r_3  
\end{pmatrix}=\frac{1}{x^2+y^2+1}
\begin{pmatrix}
	x^2+y^2 & x-iy \\
	x+iy & 1  
\end{pmatrix}.
\end{equation}
\end{definition}
It is easy to check that $\Tr(\rho^2_\mathbf{ x})=1$. Hence, $\rho_\mathbf{ x}$ always represents a
pure state for any value of the features $x$ and $y$.
\\Following the standard definition of the Bloch sphere, it can be verified that
$r_i=\Tr{(\rho_\mathbf{ x}\cdot\sigma_i)},$ with $i\in\{1,2,3\}$ and $\sigma_i$ are Pauli
matrices.

\begin{example}
Let us consider the pattern $\mathbf{ x}=(1,3)$.
The corresponding $\rho_\mathbf{ x}$ reads
$$\rho_\mathbf{ x}=\frac{1}{11}
	\begin{pmatrix}10 & 1-3i \\
          1+3i & 1  
	\end{pmatrix}.$$
\end{example}

The introduction of the density pattern leads to two different developments. The first is showed in the next Section and consists in the representation of the NMC in quantum terms. Moreover, in Section \ref{qcp}, starting from the framework of density patterns, it will be possible to introduce a Quantum Classifier that exhibits better performances than the NMC.
%%%%%%%%%%%%%%%%%%%%%%%%%%%%%%%%%%%%%%%%%%%%%%%%%%%%%%%%%%%%%%%%%%%%%%%%%%%%%%%%
\section{Classification process for density patterns}\label{sec:cp}
%%%%%%%%%%%%%%%%%%%%%%%%%%%%%%%%%%%%%%%%%%%%%%%%%%%%%%%%%%%%%%%%%%%%%%%%%%%%%%%%

As introduced in Section \ref{sec:pre}, the NMC is based on the
computation of the minimum Euclidean distance between the pattern to be
classified and the centroids of each class. In the previous Section, a quantum
counterpart of an arbitrary ``classical" pattern was provided. In order to
obtain a quantum counterpart of the standard classification process, we
need to provide a suitable definition of distance $d$ between DPs. In addition
to satisfy the standard conditions of metric, the distance $d$ also needs to
satisfy the \emph{preservation of the order}: given three arbitrary patterns
$a,b,c$ such that $d_E(a,b)\leq d_E(b,c)$, if $\rho_a,\rho_b,\rho_c$ are the DPs
related to $a,b,c$ respectively, then $d(\rho_a,\rho_b)\leq d(\rho_b,\rho_c).$
In order to fulfill all the previous conditions, we obtain the following
definition.

\begin{definition}[Normalized Trace Distance] \label{ntr}
The normalized trace distance $\dTrK$ between two arbitrary density patterns
$\rho_a$ and $\rho_b$ is given by formula 
\begin{equation}
\dTrK(\rho_a,\rho_b) = K_{a,b}\dTr(\rho_a,\rho_b),
\end{equation}
where $\dTr(\rho_a,\rho_b)$ is the standard trace distance, $\dTr(\rho_a,\rho_b)
= \frac{1}{2}\sum_i|\lambda_i|$, with $\lambda_i$ representing the eigenvalues
of $\rho_a-\rho_b$ \cite{Barnett,Nielsen}, and $K_{a,b}$ is a normalization
factor given by $K_{a,b}=\frac{2}{\sqrt{(1-r_{a_3})(1-r_{b_3})}}$, with
$r_{a_3}$ and $r_{b_3}$ representing the third Pauli components of $\rho_a$ and
$\rho_b$, respectively. 
\end{definition}

\begin{proposition}\label{ntd}
Given two arbitrary patterns $a=(x_a,y_a)$ and $b=(x_b,y_b)$ and their
respective density patterns, $\rho_a$ and $\rho_b$, we have that
\begin{equation}
\dTrK(\rho_a,\rho_b)=\dE(a,b).
\end{equation}

\begin{proof}
It can be verified that the eigenvalues of the
matrix $\rho_a-\rho_b$ are given by
\begin{equation}
Eig(\rho_a-\rho_b)=\pm\frac{\dE(a,b)}{\sqrt{(1+x_a^2+y_a^2)(1+x_b^2+y_b^2)}}.
\end{equation}
Using the definition of trace distance, we have
\begin{equation}
\Tr\sqrt{(\rho_a-\rho_b)^2}=\frac{\dE(a,b)}{\sqrt{(1+x_a^2+y_a^2)(1+x_b^2+y_b^2)}}.
\end{equation}

By applying formula (\ref{r123}) to both $r_{a_3}$ and $r_{b_3}$, we obtain that
\begin{equation}
K_{a,b}=\frac{2}{\sqrt{(1-r_{a_3})(1-r_{b_3})}}=\sqrt{(1+x_a^2+y_a^2)(1+x_b^2+y_b^2)}.
\end{equation}
\end{proof}
\end{proposition}
Using Proposition \ref{ntd}, one can see that the normalized trace distance $\dTrK$ satisfies the
standard metric conditions and the preservation of the order.

Due to the computational advantage of a quantum algorithm able to faster calculate the Euclidean distance \cite{wiebe15quantum}, the equivalence between the normalized trace distance and the Euclidean distance turns out to be potentially beneficial for the classification process we are going to introduce. 

Let us now consider two classes, $C_A$ and $C_B$, and the
respective centroids\footnote{Let us remark that, in general, $a^*$ and $b^*$ do not represent true centroids, but centroids estimated on the training set.} $a^*=(x_a,y_a)$ and $b^*=(x_b,y_b).$ 
The classification process based on NMC consists of finding the space regions given by the points closest to the first centroid $a^*$ or to the second centroid $b^*$. The patterns belonging to the first region are assigned to the class $C_A$, while patterns belonging to the second region are assigned to the class $C_B$.
The points equidistant from both the centroids represent the \emph{discriminant function} (DF), given by
\begin{equation}\label{df} 
\DF(x,y) = 2(x_a - x_b)x + 2(y_a - y_b)y + (|b^*|^2 - |a^*|^2)=0.
\end{equation}
Thus, an arbitrary pattern $c=(x,y)$ is assigned to the class $C_A$ (or $C_B$) if
$\DF(x,y) > 0$ (or $\DF(x,y) < 0$).
\\Let us notice that the Eq.~(\ref{df}) is obtained by imposing the equality
between the Euclidean distances $\dE(c,a^*)$ and $\dE(c, b^*)$. Similarly, we obtain
the quantum counterpart of the classical discriminant function.

\begin{figure}[ht!]
\centering
%\subfigure[]{\label{fig:class-gauss}
%\includegraphics[width=0.4\columnwidth]{GaussianNewVicinoBloch.jpg}}
%\subfigure[]{\label{fig:quant-gauss}
%\includegraphics[width=0.4\columnwidth]{BlochNewGaussian.jpg}
%}

\subfigure[]{\label{fig:class-moon}
\includegraphics[width=0.44\columnwidth]{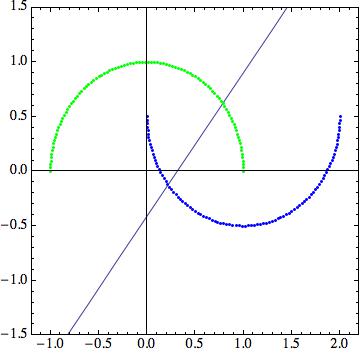}
}
\subfigure[]{\label{fig:quant-moon}
\includegraphics[width=0.48\columnwidth]{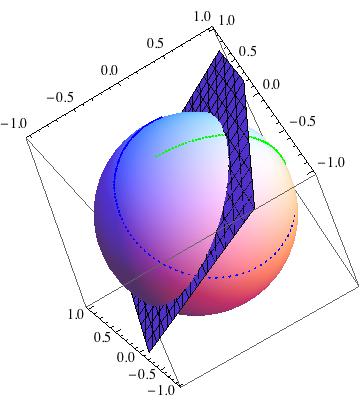}
}
\caption{Comparison between the discrimination procedures for the
\emph{Moon} dataset in $\mathbb
R^2$ \subref{fig:class-moon} and in the Bloch
sphere $\mathbb S^2$ \subref{fig:quant-moon}.}
\label{fig:discr-compare}
\end{figure}

\begin{proposition}
Let $\rho_{a^*}$ and $\rho_{b^*}$ be the DPs related to the centroids $a^*$ and
$b^*$, respectively. Then, the \emph{quantum discriminant function (QDF)} is defined as
\begin{equation}\label{qdf}
\QDF(r_1,r_2,r_3) = \vec{F}(r_{a^*},r_{b^*})^T\cdot \vec{r} + \tilde{K}^2 - 1 = 0
\end{equation}
where
\begin{itemize}
	\item $\vec{r} = (r_1, r_2, r_3)$,
	\item $\{r_{a^*_i}\}$, $\{r_{b^*_i}\}$ are Pauli components of $\rho_{a^*}$ and $\rho_{b^*}$ respectively,
	\item $\tilde{K} = \tilde{K}(r_{a^*_3}, r_{b^*_3}) = \frac{K_{c,a^*}}{K_{c,b^*}} = \sqrt{\frac{1 - r_{a^*_3}}{1 - r_{b^*_3}}},$
	\item $\vec{F}(r_{a^*},r_{b^*}) = (r_{a^*_1} - \tilde{K}^2r_{b^*_1}, r_{a^*_2} - \tilde{K}^2r_{b^*_2}, r_{a^*_3} - \tilde{K}^2r_{b^*_3}).$
\end{itemize}

\begin{proof}
In order to find the $QDF$, we use the equality between the normalized trace
distances $K_{c,a^*}\dTr(\rho_c,\rho_{a^*})$ and $K_{c,b^*}\dTr(\rho_c,\rho_{b^*})$,
where $\rho_c$ is a generic DP with Pauli components $r_1$, $r_2$,
$r_3$. We have
\begin{equation}
\begin{split}
K_{c,a^*}\dTr(\rho_c,\rho_{a^*}) &= \sqrt{\frac{(r_1 - r_{a^*_1})^2 + (r_2 - r_{a^*_2})^2 + (r_3 - r_{a^*_3})^2}{(1 - r_{a^*_3})(1 - r_3)}}, \\
K_{c,b^*}\dTr(\rho_c,\rho_{b^*}) &= \sqrt{\frac{(r_1 - r_{b^*_1})^2 + (r_2 - r_{b^*_2})^2 + (r_3 - r_{b^*_3})^2}{(1 - r_{b^*_3})(1 - r_3)}}. 
\end{split}
\end{equation}
The equality $K_{c,a^*}\dTr(\rho_c,\rho_{a^*}) = K_{c,b^*}\dTr(\rho_c,\rho_{b^*})$ reads
\begin{equation}
\sum_{i=1}^3 r_i^2 + \sum_{i=1}^3 r_{a^*_i}^2 - 2\sum_{i=1}^3 r_i r_{a^*_i} = 
\frac{1-r_{a^*_3}}{1 - r_{b^*_3}}\Big (\sum_{i=1}^3 r_i^2 + \sum_{i=1}^3 r_{b^*_i}^2 - 2\sum_{i=1}^3 r_i r_{b^*_i}\Big ).
\end{equation}
In view of the fact that $\rho_{a^*}$, $\rho_{b^*}$ and $\rho_{c}$ are pure
states, we use the conditions $\sum_{i=1}^3 r^2_{a^*_i} = \sum_{i=1}^3
r^2_{b^*_i} = \sum_{i=1}^3 r^2_i = 1$ and we get
\begin{equation}
\sum_{i=1}^3 \Big (r_{a^*_i} - \frac{1-r_{a^*_3}}{1 - r_{b^*_3}}r_{b^*_i}\Big )r_i + \frac{1-r_{a^*_3}}{1 - r_{b^*_3}} - 1 = 0.
\end{equation}
\end{proof}
\end{proposition}
This completes the proof.\\
Similarly to the classical case, we assign the DP $\rho_c$ to the class $C_A$
(or $C_B$) if $\QDF(r_1,r_2,r_3) > 0$ (or $\QDF(r_1,r_2,r_3) < 0$). Geometrically,
Eq. (\ref{qdf}) represents the surface equidistant from the DPs $\rho_{a^*}$
and $\rho_{b^*}$. 
\\Let us remark that, if we express the Pauli components $\{r_{a^*_i}\}$,
$\{r_{b^*_i}\}$ and $\{r_i\}$ in terms of classical features by
Eq.~(\ref{r123}), then Eq.~(\ref{qdf}) exactly corresponds to Eq.~(\ref{df}). As
a consequence, given an arbitrary pattern $c=(x,y)$, if $\DF(c)>0$ (or $\DF(c)<0$) then its relative DP $\rho_c$
will satisfy $\QDF(\rho_c)>0$ (or $\QDF(\rho_c)<0$, respectively). 
\\The comparison between the classical and quantum discrimination functions for
the \emph{Moon} dataset is presented in Fig.~\ref{fig:discr-compare}. Plots in
Figs.~\ref{fig:class-moon} and \ref{fig:quant-moon} present the classical and
quantum discrimination, respectively.

It is worth noting that the correspondence between pattern expressed as a feature vector (according to the standard pattern recognition approach) and pattern expressed as a density operator is quite general. Indeed, it is not related to a particular classification algorithm (NMC, in the previous case) nor to the specific metric at hand (the Euclidean one).
Therefore, it is possible to develop a similar correspondence by using other kinds of metrics and/or classification algorithms, different from NMC, adopting exactly the same approach.

This result suggests potential developements which consist in finding a quantum algorithm able to implement the normalized trace distance between density patterns. So, it would correspond to implement the NMC on a quantum computer with the consequent well known advantages.
The next Section is devoted to explore another developement, that consists in using the framework of density patterns in order to introduce a purely quantum classification process (without any classical counterpart) more convenient than the NMC on a classical computer.

\section{Quantum classification procedure}\label{qcp}
In Section \ref{sec:cp} we have shown that the NMC can be expressed by means of quantum formalism, where each pattern is replaced by a corresponding density pattern and the Euclidean distance is replaced by the normalized trace distance. Representing classical data in terms of quantum objects seems to be particularly promising in quantum machine learning. Quoting Lloyd et al.\cite{lloyd13quantum} ``Estimating distances between vectors in $N$-dimensional vector spaces takes time $O(logN)$ on a quantum computer. Sampling and estimating distances between vectors on a classical computer is apparently exponentially hard''. This convenience was already exploited in machine learning for similar tasks  \cite{wiebe15quantum,GLM}. Hence, finding a quantum algorithm for pattern classification using our proposed encoding could be particularly beneficial to speed up the classification process and it can suggest interesting developments, that, however, are beyond the scopes of this paper.

What we propose in this Section is to exhibit some explicative examples to show how, on a  classical computer, our formulation can lead to meaningful improvements with respect to the standard NMC. We also show that these improvements could be further enhanced by combining classical and quantum procedures.   
\subsection{Description of the Quantum Classifier (QC)}
In order to get a real advantage in the classification process we need to be not confined in a pure representation of the classical procedure in quantum terms. For this reason, we introduce a purely quantum representation where we consider a new definition of centroid. The basic idea is to define a \emph{quantum centroid} not as the stereographic projection of the classical centroid, but as a convex combination of density patterns.
\begin{definition} {\bf(Quantum Centroid)}
Given a dataset $\{P_1,\ldots,P_n\}$ with \\$P_i=(x_i,y_i)$ let us consider the respective set of density patterns $\{\rho_1,\ldots,\rho_n\}.$ 
The Quantum Centroid is defined as:
$$\rho_{QC}=\frac{1}{n}\sum_{i=1}^n\rho_i.$$
\end{definition}
Generally, $\rho_{QC}$ is a mixed state that has not an intuitive counterpart in the standard representation of pattern recognition, but it turns out to be convenient in the classification process. Indeed, the quantum centroid includes some further information that the classical centroid generally descards. In fact, the classical centroid does not involve all the information about the distribution of a given dataset, {\em i.e.} the classical centroid is invariant under uniform scaling transformations of the data. Consequently, the classical centroid does not take into account any dispersion phenomena. Standard pattern recognition conpensates for this lack by involving the covariance matrix \cite{DuHa}. \\On the other hand the quantum centroid is not invariant under uniform scaling. Let us consider the set of $n$ points $\{P_1,\ldots,P_n\}$ where $P_i=(x_i,y_i)$ and let $C=(c_x,c_y)=(\frac{1}{n}\sum_{j=1}^n x_j, \frac{1}{n}\sum_{j=1}^n y_j)$ be the respective classical centroid. A uniform rescaling of the $n$ points of the dataset corresponds to move each point $P_i$ along the line  joining itself with $C$, whose
generic expression is given by: $y_{x_i}=\frac{x-c_x}{x_i-c_x}(y-c_y)+c_y.$ Let  $\tilde{P}_i=(\tilde{x}_i,y_{\tilde{x}_i})$ be a generic point on this line.
Obviously, a uniform rescaling of $P_i$ by a real factor $\alpha$ is represented by the map: ${\tilde P}_i=(\tilde{x}_i,y_{\tilde {x}_i})\mapsto \alpha{\tilde P}_i=(\alpha{\tilde x}_i,y_{\alpha\tilde {x}_i}).$
Even if the classical centroid is not dependent on the rescaling factor $\alpha$, on the other hand the expression of the quantum centroid is: $$\rho_{QC}=\frac{1}{n}\begin{pmatrix}
    \sum_{i=1}^n\frac{(\alpha\tilde{x}_i)^2+(y_{\alpha\tilde{x}_i})^2}{(\alpha\tilde{x}_i)^2+(y_{\alpha\tilde{x}_i})^2+1} & \sum_{i=1}^n\frac{\alpha\tilde{x}_i-iy_{\alpha\tilde{x}_i}}{(\alpha\tilde{x}_i)^2+(y_{\alpha\tilde{x}_i})^2+1}   \\
    \sum_{i=1}^n\frac{\alpha\tilde{x}_i+iy_{\alpha\tilde{x}_i}}{(\alpha\tilde{x}_i)^2+(y_{\alpha\tilde{x}_i})^2+1}   & \sum_{i=1}^n\frac{1}{(\alpha\tilde{x}_i)^2+(y_{\alpha\tilde{x}_i})^2+1}   
 
 \end{pmatrix}$$
that, clearly, is dependent on $\alpha.$
According to the same framework used in Section \ref{sec:cp}, given two classes $C_A$ and $C_B$ of real data, let $\rho_{QCa}$ and $\rho_{QCb}$ the respective quantum centroids. Given a pattern $P$ and its respective density pattern $\rho_P$, $P$ is assigned to the class $C_A$ (or $C_B$) if $d_{tr}(\rho_P,\rho_{QCa})<d_{tr}(\rho_P,\rho_{QCb})$ (or $d_{tr}(\rho_P,\rho_{QCa})>d_{tr}(\rho_P,\rho_{QCb})$, respectively). Let us remark that we do not need any normalization parameter to be added to the trace distance $d_{tr}$, because the exact correspondence with the Euclidean distance is no more a necessary requirement in this framework. From now on we refer to the classification process based on density patterns, quantum centroids and trace distances as the \emph{Quantum Classifier} (QC).

We have shown that the quantum centroid is not independent on the dispersion of the patterns and, intuitively, it could contain some additional information with respect to the classical centroid. Consequently, it is reasonable to expect that QC could provide some better performances than the NMC. 
The next subsection will be devoted to exploit this convenience by means of numerical simulations on different datasets.    
\\Before presenting the experimental results, let briefly remark in what consists the ``convenience'' of a classification process with respect to another.
In order to evaluate the performances of a supervised learning algorithm, for each class it is tipical to refers to the respective confusion matrix \cite{F}. It is based on four possible kinds of outcome after the classification of a certain pattern: 
\begin{itemize}
\item True positive (TP): pattern correctly assigned to its class;
\item True negative (TN): pattern correctly assigned to another class;
\item False positive (FP): pattern uncorrectly assigned to its class;
\item False negative (FN): pattern uncorrectly assigned to another class.
\end{itemize} 
According to above, it is possible to recall the following definitions able to evaluate the performance of an algorithm\footnote{For the sake of the simplicity, from now on we indicate $\sum_{j=1}^C TP_j$ with TP. Similarly for TN, FP and FN.}.
\\True Positive Rate (TPR), or Sensitivity or Recall: $TPR=\frac{TP}{TP+FN}$; False Positive Rate (FPR): $FPR=\frac{FP}{FP+TN}$; True Negative Rate (TNR): $TNR=\frac{TN}{TN+FP}$; False Negative Rate (FNR): $FNR=\frac{FN}{FN+TP}$.
\\Let us consider a dataset of $C$ elements allocated in $m$ different classes. We also recall the following basic statistical notions:
\begin{itemize}
\item Error: $E=1-\frac{TP}{C};$
\item Accuracy: $Ac=\frac{TP+TN}{C};$
\item Precision: $Pr=\frac{TP}{TP+FP}.$
\end{itemize}

Further, another statistical index that is very useful to indicate the reliability of a classification process is given by the Cohen's $k$, that is $k=\frac{Pr(a)-Pr(e)}{1-Pr(e)}$, where $Pr(a)=\frac{TP+TN}{C}$ and $Pr(e)=\frac{(TP+FP)(TP+FN) + (FP+TN)(TN+FN)}{C^2}$. The value of $k$ is such that $-1\le k \le 1$, where the case $k=1$ corresponds to a perfect classification procedure.

\subsection{Implementing the Quantum Classifier}
In this subsection we implement the QC on different datasets and we show the difference between QC and NMC in terms of the values of error, accuracy, precision and other probabilistic indexes summarized above.
\\We will show how our quantum classification procedure exhibits a convenience with respect to the NMC on a classical computer by using different datasets.
\\We refer to the following very popular two-features datasets, extracted from common machine learning repositories: the \emph{Gaussian} and the \emph{Moon} datasets, composed of $200$ patterns allocated in two different classes, the \emph{Banana} dataset, composed of $5300$ patterns allocated in two classes and the \emph{3ClassGaussian}, composed of $150$ patterns allocated in three classes.
\subsubsection{Gaussian dataset}
This dataset consists of $200$ patterns allocated in two classes (with equal size), following Gaussian distributions whose means are $\mu_1 = (1,1)$, $\mu_2 = (2,2)$ and covariance matrices are $\Sigma_1 = diag(20,50)$, $\Sigma_2 = diag(5,5)$, respectively.\\
As depicted in Figure 2, the classes appear particularly mixed and the QC is able to classify a number of true positive patterns that is significantly larger than the NMC. Hence, the error of the QC is (about $20\%$) smaller than the error of the NMC. In particular, the QC turns out to be strongly beneficial in the classification of the patterns of the second class. Further, also the values related to accuracy, precision and the other statistical indexes exhibit relevant inprovements with respect to the NMC.\\
On the other hand, there are some patterns correctly classified by the NMC which are neglected by the QC.
%On this basis, it makes sense to consider a combination of both classifiers, shown in Fig 2(d), %whose performances are summarized in table 1.

On this basis, exploiting their complementarity, it makes sense to consider a combination of both classifiers. The so-called \emph{oracle} is an hypothetical selection rule that, for each pattern, is able to select the most appropriate classifier.  Its aim is to show the potentiality of an ensemble of classifiers (in this case, QC and NMC) if we were able to select the most appropriate classifier depending on the test pattern. Fig 2(d) shows the effect of the oracle whose performances are summarized in Table 1.  These performances represent the theoretical upper bound of the ensemble composed by QC and NMC.\\
We denote the variables listed in the tables as follows: E= Error; Ei= Error on the class i; Ac= Accuracy; Pr= Precision; k=Cohen's k; TPR= True positive rate; FPR=False positive rate; TNR= True negative rate; FNR= False negative rate. Let us remark that: \emph{i}) the values listed in the table are referred to the mean values over the classes; \emph{ii}) in case the number of classes is equal to $2$, is $TPR=TNR$, $FPR=FNR$ and $Ac=Pr$.

\begin{figure}[ht!]
\centering
\subfigure[]{\label{fig:gaussdataset}
\includegraphics[width=0.45\columnwidth]{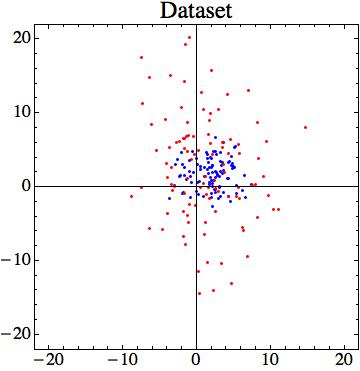}}
\subfigure[]{\label{fig:classicgaussian}
\includegraphics[width=0.45\columnwidth]{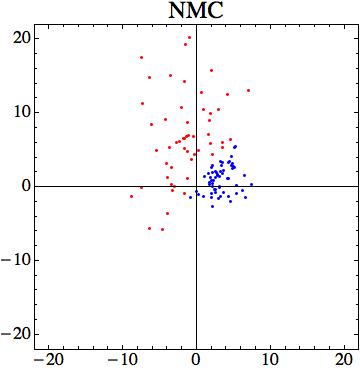}
}

\subfigure[]{\label{fig:quantumgaussian}
\includegraphics[width=0.45\columnwidth]{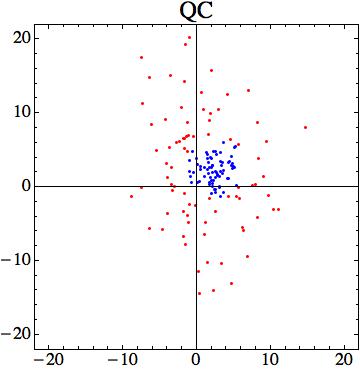}
}
\subfigure[]{\label{fig:classquantgaussian}
\includegraphics[width=0.45\columnwidth]{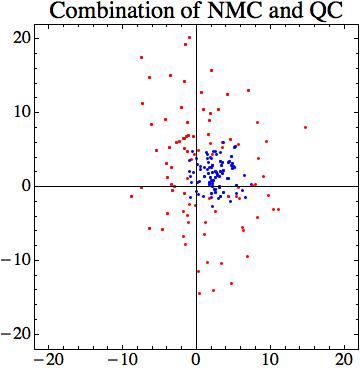}
}
\caption{Experimental results obtained fro the Gaussian dataset: (a) dataset used in the experiments, (b) classification obtained using NMC, (c) classification obtained using QC, (d) classification using the combination of NMC and QC.}
%\label{fig:discr-compare}
\end{figure}

\begin{table}[htbp]\centering \caption{Gaussian Dataset \label{GD}}
\begin{tabular}{lll c c c c c c c}\hline
&E&E1&E2&Pr&k&TPR&FPR\\ \\ \hline\hline
NMC&\bf{0.445}&0.41&0.48&\bf{0.555}&\bf{0.11}&0.555&0.445\\
QC&\bf{0.24}&0.28&0.2&\bf{0.762}&\bf{0.52}&0.76&0.24\\
NMC-QC&\bf{0.13}&0.14&0.12&\bf{0.87}&\bf{0.74}&0.87&0.13\\
\end{tabular}
\end{table}

\subsubsection{The Moon dataset}
This dataset consists of $200$ patterns equally distributed in two classes.
In this case, the correctly classified patterns of the first class are exactly the same for both classifiers but the QC turns out to be beneficial in the classification of the second class.\\
Differently from the Gaussian dataset, for this dataset the patterns correctly classified by the NMC are a proper subset of the ones correctly classified by the QC. On this basis, the QC is fully convenient with respect to the NMC and a combination of the two classifiers is useless.

\begin{figure}[ht!]
\centering
\subfigure[]{\label{fig:moondataset}
\includegraphics[width=0.45\columnwidth]{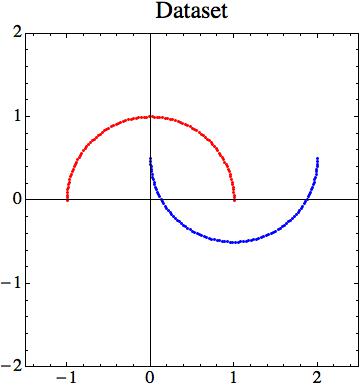}}
\subfigure[]{\label{fig:classicmoon}
\includegraphics[width=0.45\columnwidth]{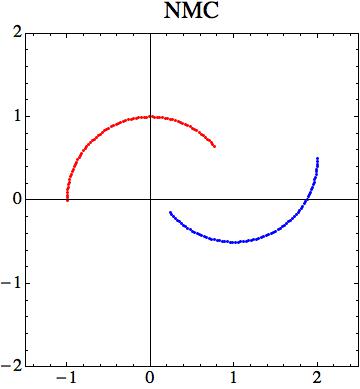}
}

\subfigure[]{\label{fig:quantummoon}
\includegraphics[width=0.45\columnwidth]{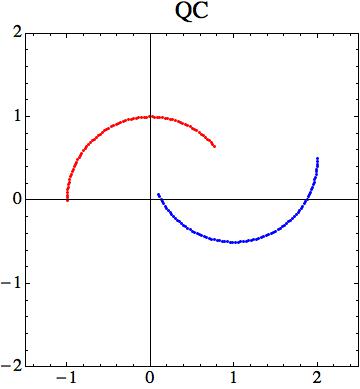}
}
%\subfigure[]{\label{fig:classquantummoon}
%\includegraphics[width=0.45\columnwidth]{ClassQuantMoon.jpg}
%}
\caption{Experimental results obtained fro the Moon dataset: (a) dataset used in the experiments, (b) classification obtained using NMC, (c) classification obtained using QC.}
%\label{fig:discr-compare}
\end{figure}

\begin{table}[htbp]\centering \caption{Moon Dataset \label{GD}}\label{M}
\begin{tabular}{ll c c c c c c c}\hline
&\bf{E}&E1&E2&\bf{Pr}&\bf{k}&TPR&FPR\\ \\ \hline\hline
NMC&\bf{0.22}&0.22&0.22&\bf{0.78}&\bf{0.56}&0.78&0.22\\
QC&\bf{0.18}&0.14&0.22&\bf{0.822}&\bf{0.64}&0.82&0.18\
\end{tabular}
\end{table}

\subsubsection{The Banana dataset}
The Banana dataset presents a particularly complex distribution that is very hard to deal with the NMC. It consists of $5300$ patterns not equally distributed between the two classes ($2376$ patterns belonging to the first class and $2924$ belonging to the second one). In this case, the QC turns out to be beneficial  in terms of all statistical indexes and for both classes.
Similarly to the gaussian case, also for the Banana dataset the NMC is able to correctly classify some points unclassified by the QC. Indeed, the contribution that the QC provides to the NMC is noticeable by the result of the combination of both classifiers, depicted in Fig. 4(d).

\begin{figure}[ht!]
\centering
\subfigure[]{\label{fig:bananadataset}
\includegraphics[width=0.45\columnwidth]{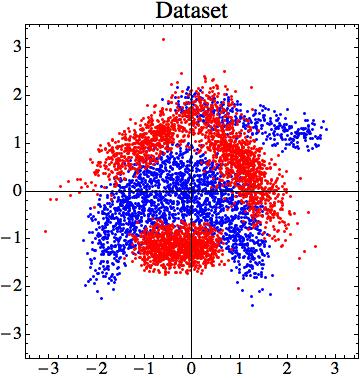}}
\subfigure[]{\label{fig:classicbanana}
\includegraphics[width=0.45\columnwidth]{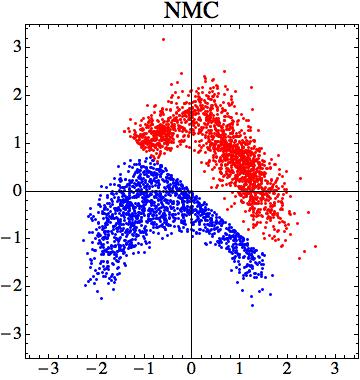}
}

\subfigure[]{\label{fig:quantumbanana}
\includegraphics[width=0.45\columnwidth]{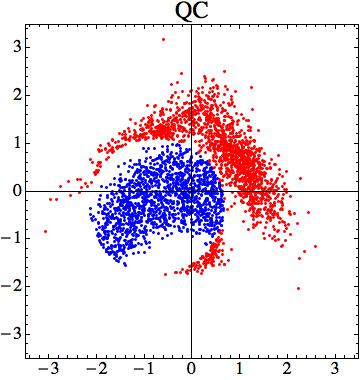}
}
\subfigure[]{\label{fig:classquantbanana}
\includegraphics[width=0.45\columnwidth]{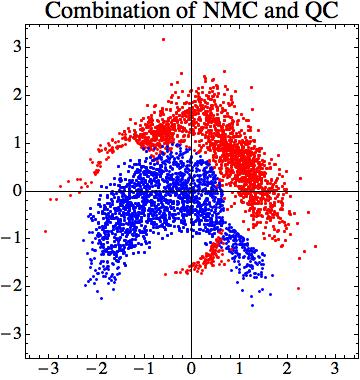}
}
\caption{Experimental results obtained fro the Banana dataset: (a) dataset used in the experiments, (b) classification obtained using NMC, (c) classification obtained using QC, (d) classification using the combination of NMC and QC.}
%\label{fig:discr-compare}
\end{figure}

\begin{table}[htbp]\centering \caption{Banana Dataset \label{GD}}\label{B}
\begin{tabular}{lll c c c c c c c}\hline
&\bf{E}&E1&E2&\bf{Pr}&\bf{k}&TPR&FPR\\ \\ \hline\hline
NMC&\bf{0.447}&0.423&0.468&\bf{0.554}&\bf{0.108}&0.555&0.445\\
QC&\bf{0.418}&0.382&0.447&\bf{0.585}&\bf{0.168}&0.585&0.415\\
NMC-QC&\bf{0.345}&0.271&0.406&\bf{0.661}&\bf{0.317}&0.662&0.338\\
\end{tabular}
\end{table}

\subsubsection{The 3ClassGaussian dataset}
In this last example we consider an equally distributed three-class dataset, consisting of 450 total number of patterns. The classes are distributed as Gaussian random variables whose means are  $\mu_1 = (-3,-3)$, $\mu_2 = (5,5)$, $\mu_3 = (7,7)$ and covariance matrices are $\Sigma_1 = diag(50,100)$, $\Sigma_2 = diag(10,5)$, $\Sigma_3 = diag(30,70)$, respectively.\\ Once again, the computation of the error and the other statistical indexes evaluated for both QC and NMC shows that the first is more convenient. Also in this case a further convenience could be reached by combining QC and NMC together. In this case the mean error decreases up to about $0.244.$

\begin{table}[htbp]\centering \caption{3Gaussian Dataset \label{3G}}
\begin{tabular}{ll c c c c c c c c c c c c c}\hline
&\bf{E}&E1&E2&E3&\bf{Ac}&\bf{Pr}&\bf{k}&TPR&FPR&TNR&FNR\\ \hline\hline
NMC&\bf{0.358}&0.367&0.433&0.273&0.762&\bf{0.653}&\bf{0.466}&0.642&0.179&0.821&0.358\\
QC&\bf{0.284}&0.287&0.307&0.26&0.81&\bf{0.724}&\bf{0.575}&0.716&0.142&0.858&0.284\\
%NMC-QC&\bf{0.244}&0.233&0.28&0.22&\bf{Ac}&\bf{Pr}&\bf{k}&TPR&FPR&TNR&FNR\\
\end{tabular}
\end{table}

\begin{figure}[ht!]
\centering
\subfigure[]{\label{fig:ThreeGaussDataset}
\includegraphics[width=0.45\columnwidth]{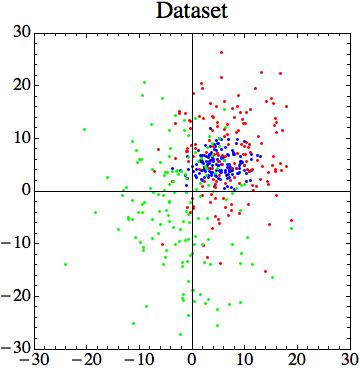}}
\subfigure[]{\label{fig:ClassicThreeGaussian}
\includegraphics[width=0.45\columnwidth]{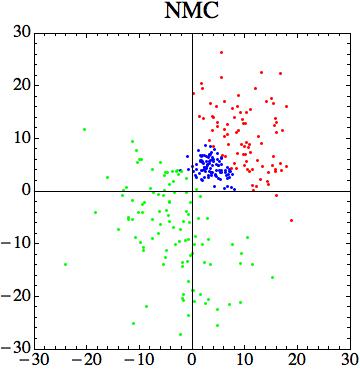}
}

\subfigure[]{\label{fig:QuantumThreeGaussian}
\includegraphics[width=0.45\columnwidth]{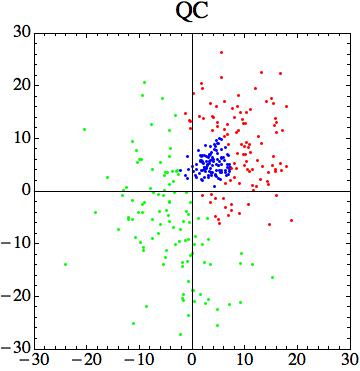}
}
\subfigure[]{\label{fig:ClassQuantThreeGauss1}
\includegraphics[width=0.45\columnwidth]{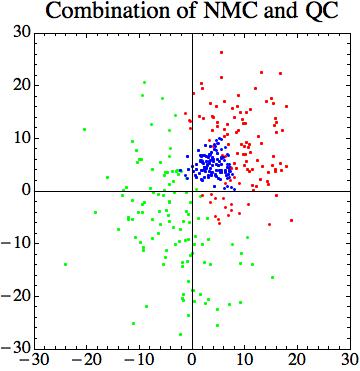}
}
\caption{Experimental results obtained fro the 3Gaussian dataset: (a) dataset used in the experiments, (b) classification obtained using NMC, (c) classification obtained using QC, (d) classification using the combination of NMC and QC.}
\label{ClassQuantThreeGauss1}
\end{figure}

\bigskip
Even if the previous examples have shown how the QC can be particularly beneficial with respect to the NMC, according to the well known \emph{No Free Lunch Theorem} \cite{DuHa}, there is no a classifier whose performance is better than the others for any dataset \cite{DuHa}. This paper is focused on the comparison between the NMC and the QC because these methods are exclusively based on the pattern-centroid distance. Anyway, a widely comparison among the QC and other commonly used classifiers (such as the LDA - Linear Discriminant Analysis - and the QDA - Quadratic Discriminant Analysis -) will be proposed for future works, where also other quantum metrics (such as the Fidelity, the Bures distance etc) instead of the trace distance will be considered to provide an adaptive version of the quantum classifier.

\section{Geometrical generalization of the model}\label{sec:general}
%%%%%%%%%%%%%%%%%%%%%%%%%%%%%%%%%%%%%%%%%%%%%%%%%%%%%%%%%%%%%%%%%%%%%%%%%%%%%%%%
In Section \ref{sec:pdp} we provided a representation of an arbitrary
two-feature pattern $\mathbf{ x}$ in the terms of a point on the surface of the Bloch
sphere $\mathbb S^2$, \ie\ a density operator $\rho_\mathbf{ x}$. A geometrical extension
of this model to the case of $n$-feature patterns inspired by quantum framework
is possible. 
\\In this section we introduce a method for representing an arbitrary
$n$-dimensional real pattern as a point in the radius-one hypersphere $\mathbb
S^n$, centered in the origin.

A quantum system described by a density operator $\rho$ in an $n$-dimensional
Hilbert space $\mathcal H$, can be represented by a linear combination of the
$n$-dimensional identity I and $2^n$ $n\times n$-square matrices $\{\sigma_i\}$
(\ie\ \emph{generalized Pauli matrices} \cite{Bertl,kimura03bloch}):
\begin{equation}\label{rn}
\rho=\frac{1}{n}I+\frac{1}{2}\sum_{i=1}^{n^2-1}r_i\sigma_i,
\end{equation}
where the real numbers $\{r_i\}$ are the Pauli components of $\rho$. Hence, by
Eq.~(\ref{rn}), a density operator $\rho$ acting on an $n$-dimensional Hilbert
space can be geometrically represented as a $(n^2-1)$-dimensional point
$P=(r_1,r_2,\ldots,r_{\tilde{n}})$ in the Bloch hypersphere $\mathbb
S^{\tilde{n}-1}$, with $\tilde{n}=n^2-1$. Therefore, by using the generalization
of the stereographic projection \cite{Karliga} we obtain the vector
$\mathbf{ x}=(x_1,x_2,\ldots,x_{\tilde{n}-1})$, that is the correspondent of $P$ in
$\mathbb R^{n^2-2}.$ In fact, the generalization of Eqs.~(\ref{sp})--(\ref{sp1}) are given by    
\begin{equation}\label{gsp}
SP_{(\tilde{n})}:(r_1,r_2,\ldots,r_{\tilde{n}})\mapsto \left (\frac{r_1}{1-r_{\tilde{n}}},\frac{r_2}{1-r_{\tilde{n}}},\ldots,\frac{r_{\tilde{n}-1}}{1-r_{\tilde{n}}}\right )=(x_1,x_2,\ldots,x_{\tilde{n}-1})
\end{equation}

\begin{align}\label{gsp1}
SP^{-1}_{(\tilde{n})}:(x_1,x_2,\ldots,x_{\tilde{n}-1})&
\mapsto\left(\frac{2x_1}{\sum_{i=1}^{\tilde n}x_i^2+1},\dots,\frac{2x_{\tilde n-1}}{\sum_{i=1}^{\tilde n}x_i^2+1},\frac{\sum_{i=1}^{\tilde{n}}x_i^2-1}{\sum_{i=1}^{\tilde n}x_i^2+1}\right)=	\nonumber\\ &=(r_1,r_2,\ldots,r_{\tilde{n}}).
\end{align}

Hence, by Eq. (\ref{gsp}), a $2$-dimensional density matrix is determined by
three Pauli components and it can be mapped onto a $2$-dimensional real vector.
Analogously, a $3$-dimensional density matrix is determined by eight Pauli
components and it can be mapped onto a $7-$dimensional real vector. Generally,
an $n$-dimensional density matrix is determined by $n^2-1$ Pauli components and
it can be mapped onto an $n^2-2$ dimensional real vector.
% The procedure to obtain the
%generalized Pauli matrices is quite standard even though it is generally not
%trivial (see Schlienz-Mahler 1995).

Now, let consider an arbitrary vector $\mathbf{ x}=(x_1, x_2,\ldots,x_{m})$ with
$(n-1)^2-1<m<n^2-2$. In this case Eq. (\ref{gsp1}) can not be applied because
$m\neq n^2-2.$ In order to represent $a$ in an $n$-dimensional Hilbert space, it
is sufficient to involve only $m+1$ Pauli components (instead of all the $n^2-1$
Pauli components of the $n$-dimensional space). Hence, we need to project the
Bloch hypersphere $\mathbb S^{n^2-2}$ onto the hypersphere $\mathbb S^m$. We
perform this projection by using Eq. (\ref{gsp1}) and by assigning some fixed
values to a number of Pauli components equal to $n^2-m-2$. In this way, we obtain a
representation in $\mathbb S^m$ that involves $m+1$ Pauli components and it
finally allows the representation of an $m$-dimensional real vector.

\begin{example}
Let us consider a vector $\mathbf{ x}=(x_1,x_2,x_3).$ By Eq. (\ref{gsp1}) we can map $\mathbf{ x}$
onto a vector $r_\mathbf{ x}=(r_1,r_2,r_3,r_4)\in\mathbb S^3.$ Hence, we need to consider
a $3$-dimensional Hilbert space $\mathcal H$. Then, an arbitrary density
operator $\rho\in\Omega_3$ can be written as
\begin{equation}
\rho=\frac{1}{3}\left(I+\sqrt3\sum_{i=1}^8r_i\sigma_i\right)
\end{equation}
with $\{r_i\}$ Pauli components such that $\sum_{i=1}^8 r_i^2\le1$ and
$\{\sigma_i\}$ generalized Pauli matrices. In this case $\{\sigma_i\}$ is the
set of eight $3\times 3$ matrices also known as \emph{Gell-Mann matrices},
namely \begin{equation}
\begin{split}
\sigma_1&=\begin{pmatrix}
    0 & 1   & 0\\
    1 & 0 & 0\\
   0 & 0   & 0
  \end{pmatrix},
\sigma_2=\begin{pmatrix}
    0 & -i   & 0\\
    i & 0 & 0\\
   0 & 0   & 0
 \end{pmatrix},
  \sigma_3=\begin{pmatrix}
    1 & 0   & 0\\
    0 & -1 & 0\\
   0 & 0   & 0
  \end{pmatrix},\\
    \sigma_4&=\begin{pmatrix}
    0 & 0   & 1\\
    0 & 0 & 0\\
   1 & 0   & 0
  \end{pmatrix},
\sigma_5=\begin{pmatrix}
    0 & 0   & -i\\
    0 & 0 & 0\\
   i & 0   & 0
 \end{pmatrix},
\sigma_6=\begin{pmatrix}
    0 & 0   & 0\\
    0 & 0 & 1\\
   0 & 1   & 0
 \end{pmatrix},\\
  \sigma_7&=\begin{pmatrix}
    0 & 0   & 0\\
    0 & 0 & -i\\
   0 & i   & 0
  \end{pmatrix},
    \sigma_8=\frac{1}{\sqrt3}\begin{pmatrix}
    1 & 0   & 0\\
    0 & 1 & 0\\
   0 & 0   & -2
 \end{pmatrix}.
  \end{split}
\end{equation}
Consequently, the generic form of a density operator $\rho$ in the
$3$-dimensional Hilbert space is given by
\begin{equation}\label{gellmann}
\rho=\frac{1}{3} \begin{pmatrix}
    \sqrt3r_3+r_8+1 & \sqrt3(r_1-ir_2)   & \sqrt3(r_4-ir_5)\\
    \sqrt3(r_1+ir_2) & -\sqrt3r_3+r_8+1 & \sqrt3(r_6-ir_7)\\
    \sqrt3(r_4+ir_5) & \sqrt3(r_6+ir_7)   & 1-2r_8
  \end{pmatrix}.
\end{equation}
Then, for any $\rho$ it is possible to associate an $8$-dimensional Bloch vector
$r=(r_1,\ldots,r_8)\in\mathbb S^7$. However, by taking $r_j=0$ for
$j=5,\ldots,8$ we obtain 
\begin{equation}
\rho_{{\bf x}}=\frac{1}{3}\begin{pmatrix}
    \sqrt3r_3+1 & \sqrt3(r_1-ir_2) & \sqrt3 r_4\\
    \sqrt3(r_1+ir_2) & -\sqrt3r_3+1 & 0\\
    \sqrt3 r_4 & 0 & 1
  \end{pmatrix}
\end{equation}
that, by Eq. (\ref{gsp1}), can be seen as point projected in $\mathbb S^3,$ where
\begin{equation}
SP^{-1}_{(4)}({\bf x})=r_{{\bf x}}=\left(\frac{2x_1}{\sum_{i=1}^3x_i^2+1},\frac{2x_2}{\sum_{i=1}^3x_i^2+1},\frac{2x_3}{\sum_{i=1}^3x_i^2+1},\frac{\sum_{i=1}^3x_i^2-1}{\sum_{i=1}^3x_i^2+1}\right).
\end{equation}
\end{example}
The generalization introduced above, allows the representation of arbitrary
patterns $\mathbf{ x}\in\mathbb R^n$ as points $\rho_\mathbf{ x}\in\mathbb S^n.$ Also the
classification procedure introduced in Section \ref{sec:cp} can be naturally
extended for an arbitrary $n$-feature pattern where the normalized trace
distance between two DPs $\rho_a$ and $\rho_b$ can be expressed using Eq.~(\ref{gsp}) in terms of the respective Pauli components as
\begin{equation}
\dTrK(\rho_a,\rho_b)=\frac{\sqrt{\sum_{i=1}^n[(r_{a_i}-r_{b_i})-(r_{a_i}r_{a_{n+1}}-r_{b_i}r_{a_{n+1}})]^2}}{(1-r_{a_{n+1}})(1-r_{b_{n+1}})}.
\end{equation}
Analogously, also the QC could be naturally extended to a $n$-dimesional problem (without lost of generality) by introducing a $n$-dimensional quantum centroid.
%%%%%%%%%%%%%%%%%%%%%%%%%%%%%%%%%%%%%%%%%%%%%%%%%%%%%%%%%%%%%%%%%%%%%%%%%%%%%%%%
\section{Conclusions and further developments}\label{sec:concl}
%%%%%%%%%%%%%%%%%%%%%%%%%%%%%%%%%%%%%%%%%%%%%%%%%%%%%%%%%%%%%%%%%%%%%%%%%%%%%%%%

In this work a quantum representation of the standard objects used in pattern recognition has been provided.
In particular, we have introduced a one-to-one correspondence between two-feature
patterns and pure density operators by using the concept of \emph{density
patterns}. 
Starting from this representation, firstly we have described the NMC in terms of quantum objects by introducing an \emph{ad hoc} definition of \emph{normalized trace distance}. We have found a quantum version of the discrimination function by means of Pauli components.  The equation of this surface was obtained by
using the normalized trace distance between density patterns and geometrically it corresponds to a surface that intersects the Bloch sphere.
This result could be considered potentially useful because suggests to find an appropriate quantum algorithm able to implement the normalized trace distance between density patterns. In this way, we could reach a replacement of the NMC in a  quantum computer, with a consequent significative reduction of the computational  complexity of the process.

Secondly, the definition of a \emph{quantum centroid} that has not any kind of classical counterpart permits to introduce a purely quantum classifier. The convenience of using this new quantum centroid lies in the fact that it seems to contain some additional information with respect to the classical one because the first takes into account also the distribution of the patterns.
The main implementative result of the paper consists in showing how the quantum classifier performs a meaningful reduction of the error and improvement of the accuracy, precision and other statistical parameters of the algorithm with respect to the NMC. Further developments will be devoted to compare our quantum classifier with other kinds of commonly used classical classifiers.

Finally, we have presented a generalization of our model that allows to express arbitrary $n$-feature patterns as points on the hypersphere
$S^n$, obtained by using the generalized stereographic projection. However, even
if it is possible to associate points of a $n$-hypersphere to $n$-feature
patterns, those points do not generally represent density operators. In
\cite{kimura03bloch,jakobczyk01geometry,kimura05bloch} the authors found some
conditions that guarantee the one-to-one correspondence between points on
particular regions of the hypersphere and density matrices. A full development
of our work is therefore intimately connected to the study on the geometrical
properties of the generalized Bloch sphere.

%\bigskip

\begin{acknowledgements} This work has been partly supported by the project  ``Computational quantum structures at the service of pattern recognition: modeling uncertainty'' [CRP-59872] funded by Regione Autonoma della Sardegna, L.R. 7/2007  (2012).
\end{acknowledgements}

%%%%%%%%%%%%%%%%%%%%%%%%%%%%%%%%%%%%%%%%%%%%%%%%%%%%%%%%%%%%%%%%%%%%%%%%%%%%%%%%
%\bibliography{density_patterns}

\begin{thebibliography}{10}

\bibitem{AeDho}
D.~Aerts, B.~D'Hooghe.
\newblock Classical logical versus quantum conceptual thought: examples in
  economics, decision theory and concept theory.
\newblock {\em Quantum interaction}, {\em Lecture Notes in Comput. Sci.}, {\bf 5494}: 128--142. Springer, Berlin (2009).

\bibitem{AGS}
D.~Aerts, L.~Gabora, S.~Sozzo.
\newblock Concepts and their dynamics: A quantum-theoretic modeling of human
  thought.
\newblock {\em Topics in Cognitive Science}, {\bf5}(4): 737--772 (2013).

\bibitem{AKN}
D.~Aharonov, A.~Kitaev, N.~Nisan.
\newblock Quantum circuits with mixed states.
\newblock In {\em Proceedings of the 30th Annual ACM Symposium on Theory of
  Computing}, 20--30. ACM (1998).

\bibitem{Barnett}
S.M. Barnett.
\newblock {\em Quantum information}, {\bf16}, {\em Oxford Master Series in
  Physics}.
\newblock Oxford University Press, Oxford.
\newblock Oxford Master Series in Atomic, Optical, and Laser Physics (2009).

\bibitem{beltrametti2014quantum-p2}
E.~Beltrametti, M.~L. Dalla~Chiara, R.~Giuntini, R.~Leporini,
  G.~Sergioli.
\newblock A quantum computational semantics for epistemic logical operators.
  part ii: Semantics.
\newblock {\em Int. J. Theor. Phys.}, {\bf53(10)}: 3293--3307 (2014).

\bibitem{beltrametti2014quantum-p1}
E.~Beltrametti, M.L. Dalla~Chiara, R.~Giuntini, R.~Leporini, G.~Sergioli.
\newblock A quantum computational semantics for epistemic logical operators.
  part i: epistemic structures.
\newblock {\em Int. J. Theor. Phys.}, {\bf53}(10): 3279--3292 (2014).

\bibitem{BenShor}
C.H. Bennett, P.W. Shor.
\newblock Quantum information theory.
\newblock {\em IEEE Trans. Inform. Theory}, 1998.

\bibitem{Bertl}
R.A. Bertlmann, P.~Krammer.
\newblock Bloch vectors for qudits.
\newblock {\em J. Phys. A}, {\bf41}(23):235303, 21 (2008).

%\bibitem{Bez} J.C.~Bezdek 
%\newblock {\em Pattern Recognition with Fuzzy Objective Function Algorithms}. 
%\newblock Boston, MA: Springer US (1981).

\bibitem{caraiman2012image}
S.~Caraiman and V.~Manta.
\newblock Image processing using quantum computing.
\newblock In {\em System Theory, Control and Computing (ICSTCC), 2012 16th
  International Conference on},  1--6, IEEE (2012).

\bibitem{chefles00quantum}
A.~Chefles.
\newblock Quantum state discrimination.
\newblock {\em Contemp. Phys.}, {\bf41}(6):401--424,
\newblock arXiv:quant-ph/0010114 (2000).

\bibitem{Cox}
H.S.M. Coxeter.
\newblock {\em Introduction to geometry}.
\newblock John Wiley \& Sons, Inc., New York-London-Sydney, 2nd edition (1969).

\bibitem{DGG}
M.L. Dalla~Chiara, R.~Giuntini, R.~Greechie.
\newblock {\em Reasoning in quantum theory: sharp and unsharp quantum logics},
  volume~22.
\newblock Springer Science \& Business Media (2004).

%\bibitem{Did} L.~Didaci, G.~Marcialis, F.~Roli.
%\newblock Analysis of unsupervised template update in biometric recognition systems.
%\newblock {\em Pattern Recognition Letters}, {\bf 37}(1):151--160 (2014).

\bibitem{DuHa} R.O.~Duda, P.E.~Hart, D.G.~Stork.
\newblock {\em Pattern Classification}.
\newblock Wiley Interscience, 2nd edition (2000).

%\bibitem{Duin} R.P.W~Duin, F.~Roli, D.~de~Ridder. 
%\newblock A note on core research issues for statistical pattern recognition.
%\newblock{\em Pattern recognition letters}, {\bf 23}(4):493--499 (2002).

\bibitem{EWL}
J.~Eisert, M.~Wilkens, and M.~Lewenstein.
\newblock Quantum games and quantum strategies.
\newblock {\em Phys. Rev. Lett.}, {\bf83}(15):3077 (1999).

\bibitem{Eldar}
Y.C. Eldar and A.V. Oppenheim.
\newblock Quantum signal processing.
\newblock {\em Signal Processing Magazine, IEEE}, {\bf19}(6):12--32 (2002).

\bibitem{F}
T. Fawcet
\newblock An Introduction to ROC Analysis
\newblock {\em Pattern Recognition Letters}, {\bf 27} (8): 861--874 (2006).

\bibitem{FSA}
H.~Freytes, G.~Sergioli, and A.~Aric{\`o}.
\newblock Representing continuous {$t$}-norms in quantum computation with mixed
  states.
\newblock {\em J. Phys. A}, {\bf43}(46):465306, 12 (2010).

\bibitem{GLM} V.~Giovannetti, S.~Lloyd, L.~Maccone.
\newblock Quantum random access memory.
\newblock {\em Phys. Rev. L}, {\bf100}(16):160501, (2008).

%\bibitem{Gor} J.~Gordon, E.H.~Shortliffe.
%\newblock {\em The Dempster-Shafer theory of evidence}. 
%\newblock Morgan Kaufmann Publishers Inc. San Francisco, CA, USA, 529–-539 (1990).

\bibitem{HaTi}
T.~Hastie, R.~Tibshirani, J.~Friedman.
\newblock \emph{The Elements of Statistical Learning}. 
\newblock Springer (2001).


\bibitem{hayashi05quantum}
A.~Hayashi, M.~Horibe, and T.~Hashimoto.
\newblock Quantum pure-state identification.
\newblock {\em Phys. Rev. A}, {\bf72}(5):052306 (2005).

\bibitem{helstrom}
C.W. Helstrom.
\newblock {\em Quantum detection and estimation theory}.
\newblock Academic Press (1976).

\bibitem{Jaeg}
G.~Jaeger.
\newblock {\em Quantum information}.
\newblock Springer, New York,
\newblock An overview, With a foreword by Tommaso Toffoli (2007).

\bibitem{Jaeg2}
G.~Jaeger.
\newblock {\em Entanglement, information, and the interpretation of quantum
  mechanics}.
\newblock Frontiers Collection. Springer-Verlag, Berlin (2009).

\bibitem{jakobczyk01geometry}
L.~Jak{\'o}bczyk and M.~Siennicki.
\newblock Geometry of bloch vectors in two-qubit system.
\newblock {\em Phys. Lett. A}, {\bf286}(6):383--390 (2001).

%\bibitem{kandel1999introduction}
%A.~Kandel.
%\newblock {\em Introduction to pattern recognition: statistical, structural,
%  neural, and fuzzy logic approaches}, {\bf32},
%\newblock World Scientific (1999).

\bibitem{Karliga}
B.~Karl{\i}{\v{g}}a.
\newblock On the generalized stereographic projection.
\newblock {\em Beitr\"age Algebra Geom.}, {\bf37}(2):329--336 (1996).

\bibitem{kimura03bloch}
G.~Kimura.
\newblock The {Bloch} vector for {N}-level systems.
\newblock {\em Phys. Lett. A}, {\bf314}(5–6):339--349 (2003).

\bibitem{kimura05bloch}
G.~Kimura and A.~Kossakowski.
\newblock The {Bloch}-vector space for {N}-level systems: the
  spherical-coordinate point of view.
\newblock {\em Open Systems \& Information Dynamics}, {\bf12}(03):207--229,
\newblock arXiv:quant-ph/0408014 (2005).

\bibitem{kolossa2011robust}
D.~Kolossa and R.~Haeb-Umbach.
\newblock {\em Robust speech recognition of uncertain or missing data: theory
  and applications}.
\newblock Springer Science \& Business Media (2011).

\bibitem{lloyd13quantum}
S.~Lloyd, M.~Mohseni, and P.~Rebentrost.
\newblock Quantum algorithms for supervised and unsupervised machine learning.
\newblock arXiv:1307.0411 (2013).

\bibitem{lloyd2014quantum}
S.~Lloyd, M.~Mohseni, and P.~Rebentrost.
\newblock Quantum principal component analysis.
\newblock {\em Nature Physics}, {\bf10}(9):631--633 (2014).

\bibitem{lu2014quantum}
S.~Lu, S.L. Braunstein.
\newblock Quantum decision tree classifier.
\newblock {\em Quantum Inf. Process.}, {\bf13}(3):757--770 (2014).

\bibitem{manju14applications}
A.~Manju, M.J. Nigam.
\newblock Applications of quantum inspired computational intelligence: a
  survey.
\newblock {\em Artificial Intelligence Review}, {\bf42}(1):79--156 (2014).

\bibitem{manning2008introduction}
C.D Manning, P.~Raghavan, and H.~Sch{\"u}tze.
\newblock {\em Introduction to information retrieval}, volume~1.
\newblock Cambridge university press Cambridge (2008).

\bibitem{miszczak12high-level}
J.A. Miszczak.
\newblock {\em High-level Structures for Quantum Computing}, {\bf6} {\em
  Synthesis Lectures on Quantum Computing}.
\newblock Morgan \& Claypool Publishers (2012).

\bibitem{N}
E.~Nagel.
\newblock Assumptions in economic theory.
\newblock {\em The American Economic Review}, 211--219 (1963).

\bibitem{Nielsen}
M.A. Nielsen and I.L. Chuang.
\newblock {\em Quantum computation and quantum information}.
\newblock Cambridge University Press, Cambridge (2000).

\bibitem{Ohya}
M.~Ohya and I.~Volovich.
\newblock {\em Mathematical foundations of quantum information and computation
  and its applications to nano- and bio-systems}.
\newblock Theoretical and Mathematical Physics. Springer, Dordrecht (2011).

\bibitem{ostaszewski15quantum}
M.~Ostaszewski, P.~Sadowski, P.~Gawron.
\newblock Quantum image classification using principal component analysis.
\newblock {\em Theoretical and Applied Informatics}, {\bf27}:3
\newblock arXiv:1504.00580 (2015).

%\bibitem{Ped} W.~Pedrycz.
%\newblock Fuzzy sets in pattern recognition: Methodology and methods. 
%\newblock {\em Pattern Recognition}, {\bf 23}(1): 121–-146 (1990).

%\bibitem{Peters}
%%N.A. Peters, T.-C. Wei, P.G. Kwiat.
%\newblock Mixed-state sensitivity of several quantum-information benchmarks.
%\newblock {\em Phys. Rev. A}, {\bf70}(5):052309 (2004).

\bibitem{sause2013quantification}
M.G.R. Sause, S.~Horn.
\newblock Quantification of the uncertainty of pattern recognition approaches
  applied to acoustic emission signals.
\newblock {\em Journal of Nondestructive Evaluation}, {\bf32}(3):242--255 (2013).

\bibitem{schuld14introduction}
M.~Schuld, I.~Sinayskiy, F.~Petruccione.
\newblock An introduction to quantum machine learning.
\newblock {\em Contemp. Phys.}, {\bf56}(2),
\newblock arXiv:1409.3097 (2014).

\bibitem{schuld14quantum}
M.~Schuld, I.~Sinayskiy, F.~Petruccione.
\newblock Quantum computing for pattern classification.
\newblock In {\em PRICAI 2014: Trends in Artificial Intelligence}, 208--220, Springer (2014).

\bibitem{schuld14quest}
M.~Schuld, I.~Sinayskiy, F.~Petruccione.
\newblock The quest for a {Q}uantum {N}eural {N}etwork.
\newblock {\em Quantum Inf. Process.}, {\bf13}(11):2567--2586 (2014).

\bibitem{Schwartz}
J.M. Schwartz, H.P. Stapp, M.~Beauregard.
\newblock Quantum physics in neuroscience and psychology: a neurophysical model
  of mind-brain interaction.
\newblock {\em Philosophical Transactions of the Royal Society B: Biological
  Sciences}, {\bf360}(1458):1309--1327 (2005).

%\bibitem{Sha} G.~Shafer.
%\newblock{\em A Mathematical Theory of Evidence}.
%\newblock Princeton University Press (1976).

\bibitem{Shann}
C.E. Shannon.
\newblock A mathematical theory of communication.
\newblock {\em Bell System Tech. J.}, {\bf27}:379--423, 623--656 (1948).

\bibitem{Stapp}
H.P. Stapp.
\newblock {\em Mind, matter, and quantum mechanics}.
\newblock Springer-Verlag, Berlin (1993).

\bibitem{tanaka08quantum}
K.~Tanaka, K.~Tsuda.
\newblock A quantum-statistical-mechanical extension of gaussian mixture model.
\newblock {\em Journal of Physics: Conference Series}, {\bf95}(1):012023 (2008).

\bibitem{trugenberger2002quantum}
C.A. Trugenberger.
\newblock Quantum pattern recognition.
\newblock {\em Quantum Inf. Process.}, {\bf1}(6):471--493 (2002).

\bibitem{webb}
A.R. Webb, K.D. Copsey.
\newblock {\em Statistical Pattern Recognition}.
\newblock Wiley, 3rd edition (2011).

\bibitem{wiebe15quantum}
N.~Wiebe, A.~Kapoor, K.M. Svore.
\newblock Quantum nearest-neighbor algorithms for machine learning.
\newblock {\em Quantum Inform. Comput.}, {\bf15}(34):0318--0358 (2015).

\bibitem{Wilde}
M.M. Wilde.
\newblock {\em Quantum information theory}.
\newblock Cambridge University Press, Cambridge (2013).

\bibitem{wittek14quantum}
P.~Wittek.
\newblock {\em Quantum Machine Learning: What Quantum Computing Means to Data
  Mining}.
\newblock Academic Press (2014).

%\bibitem{Xu} L.~Xu, E.~Hung.
%\newblock Improving classification accuracy on uncertain data by considering multiple subclasses.
%\newblock {\em Neurocomputing}, {\bf 145}: 98–-107 (2014).

%\bibitem{Yag} R.R.~Yager, L.~Liu
%\newblock {\em Classic Works of the Dempster-Shafer Theory of Belief Functions}, {\bf 219}, Berlin, Heidelberg: Springer Berlin Heidelberg (2008).


\end{thebibliography}
%\bibliographystyle{spphys}
%%%%%%%%%%%%%%%%%%%%%%%%%%%%%%%%%%%%%%%%%%%%%%%%%%%%%%%%%%%%%%%%%%%%%%%%%%%%%%%%

\end{document}